\documentclass[11pt]{article}
\usepackage{bm} 
\usepackage{amsmath,amsthm}
\newtheorem{theorem}{Theorem}

\usepackage[colorinlistoftodos]{todonotes}

\usepackage{ulem}

\usepackage{color, soul}

\makeatletter
\begin{document}
\title{Statistical Neurodynamics of Deep Networks:
\\ Geometry of
Signal Spaces} \author{Shun-ichi Amari\thanks{RIKEN CBS, Wako-shi,
Japan} $^{\dag}$ \and Ryo Karakida\thanks{AIST, Tokyo,Japan} \and Masafumi
Oizumi\thanks{Araya Co., Tokyo, Japan}} \date{}

\maketitle

\begin{abstract}
Statistical neurodynamics studies macroscopic behaviors of randomly connected neural networks.  We consider a deep layered feedforward
network where input signals are processed layer by layer.  The manifold of input signals is embedded in a higher dimensional manifold of the next layer as a curved submanifold, provided the number of neurons is larger than that of inputs.  We show geometrical features of the embedded manifold, proving that the manifold enlarges or shrinks locally isotropically so that it is always embedded conformally.  We study the curvature of the embedded manifold.  The scalar curvature converges to a constant or diverges to infinity slowly.  The distance between two signals also changes, converging eventually to a stable fixed value, provided both the number of neurons in a layer and the number of layers tend to infinity.  This causes a problem, since when we consider a curve in the input space, it is mapped as a continuous curve of fractal nature, but our theory contradictorily suggests that the curve eventually converges to a discrete set of equally spaced points. In reality, the numbers of neurons and layers are finite and thus, it is expected that the finite size effect causes the discrepancies between our theory and reality. We need to further study the discrepancies to understand their implications on information processing.
\end{abstract}

\section{Introduction}

Statistical dynamics bases studies of the behaviors of macroscopic quantities such as temperature and entropy, on the microscopic physical laws of molecular particles, by using the statistical averages of random variables. Statistical neurodynamics infers the macroscopic behaviors of randomly connected neural networks from the microscopic signal processing of component neurons.  This idea originated from the work of Rozonoer (1969) and Amari (1970, 1971), and its mathematical foundation was further studied by Amari (1974) and Amari et al. (1977).

Poole et al. (2016) used the method of statistical neurodynamics to elucidate the behaviors of randomly connected deep feedforward neural networks, giving a new perspective to examine deep networks, and Schoenholz et al. (2016) applied it to backpropagation learning in deep networks.  Inspired by these studies, the present paper investigates how geometrical structures of signal spaces develop layer by layer (See also our accompanying paper (Amari, Karakida \& Oizumi, 2018) for the analysis of the Fisher information and natural gradient).

An input space is embedded in the output space of a layer when the number of neurons is larger than the number of inputs.  The embedded manifold is curved and enlarges or shrinks, eventually giving rise to a fractal structure.  The metric of a signal space develops conformally so that the tangent spaces are rotated with isotropical enlargement or shrinkage.  We study the metric and curvature based on the mean field approximation.

There are three main findings in the present paper. First, we show that the metric tensor is conformally changed through layers. Second, we explicitly calculate how the curvature (extrinsic curvature tensor and affine connection of the embedded manifold) changes through layers. It converges to a fixed value under a certain condition (chaotic regime of Poole et al., 2016) and diverges very slowly under the other condition. Third, we elucidate the distance law, which describes how the distance between two signal points changes through layers. We demonstrate that there is a contradiction between the real situation where both the number of layers and the number of neurons are finite and the ideal theoretical situation where both tend to infinity, as described below.

The distance between two signals changes as signals are processed in layers and its dynamics has a stable equilibrium point.  Such a dynamics of distance was proposed in Amari (1974) and Amari et al. (1977) and applied to randomly connected recurrent networks to study microscopic characteristics of their attractors (Amari et al. 2013; Toyoizumi et al. 2015).  Poole et al. (2016) expressed the distance law for deep layered feedforward networks in terms of the overlap of two inputs, which is equivalent to the distance between them.  However, the dynamic law of distance is problematic because the distance between any two  signals eventually converges to a constant in the limit when the number of neurons in each layer is infinitely large and the number of layers is also infinitely large.

Let us consider a curve ${\bm{x}}(s)$ in the input space, where $s$ is the parameter describing the curve.  It is embedded in a higher dimensional manifold continuously, so it is impossible that the curve converges to a set of equally spaced points in the case that both the numbers of neurons and the number layers are finite.  The curve has a fractal structure and frustrations take place between continuity and discreteness. We will finally remark the importance of the effect of the finiteness of the numbers of neurons and layers.

\section{Layered feed-forward network of random connections}

Let us consider a feed-forward network consisting of $m$ neurons that receive $n$-dimensional inputs ${\bm{x}}= \left( x_1, \cdots, x_n
\right)$ (see Figure 1). 
\begin{figure}
 \centering
 \includegraphics[width=7cm]{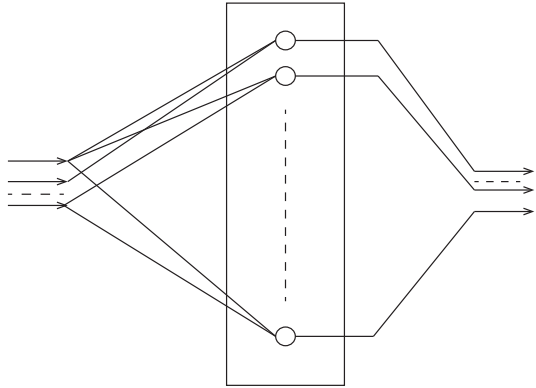}
 \caption{Input-output relation of a layer}
\end{figure} 
Let ${\bm{y}}= \left(y_1, \cdots, y_m \right)$ be the
outputs from the $m$ neurons.  Let $w_{i \kappa}$ be the connection weight from the $\kappa$-th input to neuron $i$ and $b_i$ be the bias
term of neuron $i$.  We use indices $\kappa, \lambda, \cdots$ for input
vectors and indices $i, j, \cdots$ for output vectors. 
The sum of stimuli for
neuron $i$ is
\begin{equation}
 u_i = \sum_{\kappa} w_{i \kappa} x_{\kappa} + b_i
\end{equation}
and the output from that neuron is
\begin{equation}
 y_i = \varphi \left(u_i \right),
\end{equation}
where $\varphi$ is an output function (see Figure 2). 
\begin{figure}
 \centering
 \includegraphics[width=7cm]{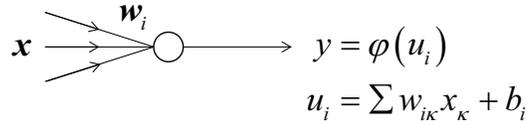}
 \caption{Input-output relation of neuron $i$}
\end{figure} 
We use the error
function
\begin{equation}
 \varphi (u) = \frac 1{\sqrt{2 \pi}} \int^u_{-\infty} \exp
 \left\{ -\frac{v^2}2 \right\} dv,
\end{equation}
because this is convenient for obtaining analytical formulas. 
The behaviors of the network are qualitatively the same if other output
functions are used.

We assume that the connection weights $w_{i \kappa}$ and the bias $b_i$
are independent random variables having Gaussian distributions with
variances $\sigma^2/n$ and $\sigma^2_b$, respectively, and mean 0.
Then, all $u_i$ are independent random variables subject to the same
Gaussian distribution with mean 0 and variance
\begin{equation}
 \tau^2 = \frac{\sigma^2}n \sum 
  x^2_{\kappa} + \sigma^2_b.
\end{equation}
The outputs $y_i= \varphi \left(u_i \right)$ are nonlinear functions of
$u_i$, so they are also independently and identically distributed.

We consider a deep network consisting of concatenated feedforward
layers.  The $t$-th layer receives input ${\bm{x}}^{t-1}$ which is the
output of the previous layer $t-1$ and emits output ${\bm{x}}^t$
(see Figure 3).  
\begin{figure}
 \centering
 \includegraphics[width=7cm]{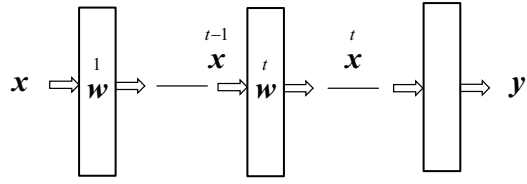}
 \caption{Deep layered network}
\end{figure} 
The connection weights and biases are $w^t_{ij}$ and $b^t_i$,
subject to the same 0 mean independent Gaussian distributions with
variance $\sigma^2_t/n_{t-1}$, and $\sigma^2_b$, respectively, where
$n^t$ is the number of neurons in layer $t$.  The output is written as
\begin{align}
 \label{eq::am520180601}
 x^t_i &= \varphi \left(u^t_i \right), \\
 \label{eq::am620180601}
 u^t_i &= \sum w^t_{ij} x^{t-1}_j + b^t_i
\end{align}
or, in the vector-matrix notation,
\begin{equation}
 \label{eq:am720180613}
 {\bm{x}}^t = \varphi \left({\bm{W}}^t {\bm{x}}^{t-1}
  + {\bm{b}}^t \right).
\end{equation}

Let $X^{t-1}$ be the input signal space consisting of
$n^{t-1}$-dimensional input signals ${\bm{x}}^{t-1}$ and $X^t$ be the
$n^t$-dimensional output signal space of layer $t$.  When $n^t \ge
n^{t-1}$, the manifold $X^{t-1}$ is embedded in $X^t$.  When
$n^t<n^{t-1}$, $X^{t-1}$ is mapped to a lower-dimensional $X^t$ by a
many-to-one mapping.

Since the original input manifold $X=X^0$ is transformed layer by layer
successively, its image at layer $t$ is denoted by $\tilde{X}^t$ which
is a submanifold embedded in $X^t$.  We infer the geometry of
$\tilde{X}^t$ from that of $\tilde{X}^{t-1}$.

\section{Propagation of activities}

We consider as a simple macroscopic quantity the activity of output
${\bm{x}}^t$ of layer $t$ defined by
\begin{equation}
 A^t = \frac 1{n^t} \sum \left(x^t_i \right)^2.
\end{equation}
This is written as
\begin{equation}
 A^t = \frac 1{n^t} \sum \left\{ \varphi \left(u^t_i \right) \right\}^2.
\end{equation}
Since $u^t_i$ are independent random Gaussian variables with mean 0 and variance
\begin{equation}
  \tau^2_t = \sigma^2_t A^{t-1} + \sigma^2_b, 
\end{equation}
the law of large numbers guarantees that it converges to the expectation
\begin{equation}
 \label{eq:am10}
 A^t = {\mathrm{E}} \left[ \left\{ \varphi \left(u^t_i \right)\right\}^2 \right],
\end{equation}
when $n^t$ is sufficiently large, where ${\mathrm{E}}$ denotes the
expectation with respect to $w^t_{ij}$ and $b^t_i$.

In order to calculate the expectations of various quantities like
$A^t$, we define a fundamental function $\chi_0$ by
\begin{equation}
 \chi_0 (\sigma^2, A) = \int \varphi^2 \left(\sigma_A v \right) D
  v, 
\end{equation}
where $\sigma^2_A = \sigma^2 A + \sigma^2_b$ and
\begin{equation}
 D v = \frac 1{\sqrt{2 \pi}} \exp \left\{ -\frac{v^2}2 \right\} dv.
\end{equation}
For later use, we define the $p$-th fundamental functions $\chi_p (p=1,
2)$ by
\begin{eqnarray}
 \chi_1 \left(\sigma^2, A \right) &=& \sigma^2 \int \left\{ \varphi'
	 \left(\sigma_A v \right)\right\}^2 Dv, \\
 \chi_2 \left(\sigma^2, A \right) &=& \sigma^4 \int \left\{ \varphi''
	 \left(\sigma_A v \right)\right\}^2 Dv.
\end{eqnarray}
Since $u^t_i$ is subject to $N \left(0, \tau^2_t \right)$, we have
\begin{align}
 {\mathrm{E}} \left[ \left\{\varphi \left(u^t_i \right)\right\}^2 \right] &=
  \frac 1{\sqrt{2 \pi} \tau_t} 
  \int \left\{\varphi(u)\right\}^2 \exp \left\{-\frac{u^2}{2
 \tau^2_t}\right\} du \\
 &= \chi_0 \left( \sigma^2_t, A^{t-1}\right).
\end{align}
When $\varphi$ is the error function,
\begin{equation}
 \chi_0 (\sigma^2, A) = \frac 1{2 \pi} \cos^{-1}
 \left( \frac{-\sigma^2_A}{1+\sigma^2_A}\right),
\end{equation}
as is seen in Appendix I.  Hence we have the following theorem,
describing how the activity develops.

\begin{theorem}\upshape
The activity develops as
\begin{equation}
 A^t = \chi_0 \left( \sigma^2_t, A^{t-1}\right).
\end{equation}

Since $\chi_0(\sigma^2, A)$ is a monotonically increasing function of
 $A$, when $\sigma^2_t=\sigma^2$ for all $t$, there is an equilibrium
 $\bar{A}$ satisfying
\begin{equation}
 \chi_0 \left(\sigma^2, \bar{A}\right) = \bar{A},
\end{equation}
that is uniquely determined and is stable (see Figure 4).  
\begin{figure}
 \centering
 \includegraphics[width=7cm]{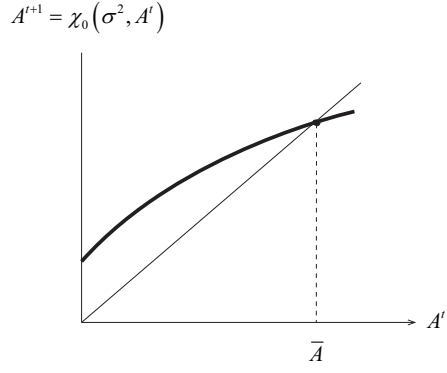}
 \caption{Dynamics of $A^t$ and its equilibrium $\bar{A}$}
\end{figure} 
In this case, as
$t$ becomes large, $A^t$ converges to $\bar{A}$ quickly.  Hence
$\tilde{X}^t$ is concentrated on the sphere of radius $\bar{A}$ in $X^t$
\begin{equation}
 \frac 1{n_t} \sum \left({x}^t_i \right)^2 = \bar{A},
\end{equation}
with small fluctuations in the radius directions.  When
$\sigma^2_t$ are different, however, the radius $A^t$ of the sphere fluctuates.
\end{theorem}

\section{Development of metric}

We consider again layer $t$, in which inputs are ${\bm{x}}^{t-1}$
belonging to $\tilde{X}^{t-1}$ and the outputs $\tilde{\bm{x}}^t$
constitute $\tilde{X}^t$.  This section treats the case of $n^t \ge
n^{t-1}$ for all $t$, so $n_0$-dimensional $\tilde{X}^{t-1}$ is embedded
in an $n_t$-dimensional manifold $X^t$.  The image $\tilde{X}^t \subset
X^t$ is a curved $n_0$-dimensional manifold, provided $n^t \ge n^{t-1}
\cdots \ge n^0=n$.

We first consider the input signal space $X=X^0$, which is assumed to be
a Euclidean space.  Let $\left\{{\bm{e}}_{\kappa}\right\}$ be the set of
orthonormal basis vectors in the input manifold $X$ along the coordinate
axis $x_{\kappa}$.  So the small line element $d{\bm{x}}$ connecting
${\bm{x}}$ and ${\bm{x}}+d{\bm{x}}$ is written as
\begin{equation}
 d{\bm{x}} = \sum d x_{\kappa}{\bm{e}}_{\kappa}.
\end{equation}

We consider the output signal submanifold $\tilde{X}^t$ of layer $t$.
Coordinate lines $x_{\kappa}$ in input $X$ become curved lines
$\tilde{x}^t_{\kappa}$ in $\tilde{X}^t$ and let
$\left\{\tilde{\bm{e}}^t_{\kappa}\right\}$ be the basis vectors in the
tangent space of $\tilde{X}^t$ along the coordinate curves $x_{\kappa}$.
See Fig. 5.
\begin{figure}
 \centering
 \includegraphics[width=7cm]{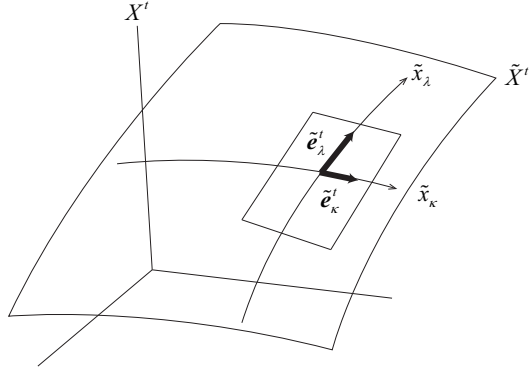}
 \caption{Tangent Space of $\tilde{X}^t$ and basis vectors  }
\end{figure} 

A small line element $d \tilde{\bm{x}}^{t-1}$ in $\tilde{X}^{t-1}$
changes to $d \tilde{\bm{x}}^t$ as
\begin{equation}
  d \tilde{x}^t_i = \sum 
   B^t_{ij} d \tilde{x}^{t-1}_j,
\end{equation}
where the Jacobian matrix ${\bm{B}}^t= \left(B^t_{i j}\right)$ of
(\ref{eq:am720180613}) is, 
\begin{equation}
B^t_{i j}= \frac{\partial \tilde{x}^t_i}{\partial \tilde{x}^{t-1}_j} =
  \frac{\partial \varphi \left(u^t_i\right)}{\partial \tilde{x}^{t-1}_j}.
\end{equation}

Both $d \tilde{\bm{x}}^t$ and $d \tilde{\bm{x}}^{t-1}$ belong to the
tangent spaces of $\tilde{X}_t$ and $\tilde{X}_{t-1}$, respectively.
The basis vector $\tilde{{\bm{e}}}^t_{\kappa}$ of the tangent space of
$\tilde{X}^t$ is a vector in $X^t$,
\begin{equation}
 \tilde{\bm{e}}^t_{\kappa} =
 \frac{\partial \tilde{\bm{x}}^t}{\partial x_{\kappa}},
\end{equation}
the $i$-th component of which is
\begin{equation}
 \left(\tilde{\bm{e}}^t_{\kappa}\right)_i =
 \frac{\partial \tilde{x}^t_i}{\partial x_{\kappa}}.
  \quad i=1, \cdots, n_t.
\end{equation}
We see that $\tilde{\bm{e}}^{t-1}_{\kappa}= \left(e^{t-1}_{\kappa j}\right)$
is mapped to $\tilde{\bm{e}}^t_{\kappa} = \left(e^{t}_{\kappa i}\right)$ linearly as
\begin{equation}
 \label{eq:am2020171124}
 e^{t}_{\kappa i} = \sum_j B^t_{ij} e^{t-1}_{\kappa j}.
\end{equation}

We introduce a metric tensor to the input signal manifold $X$ by using
the basis vectors of the $t$-th layer given by
\begin{equation}
 \label{eq:am3120180607}
 g^t_{\kappa \lambda} = \langle \tilde{\bm{e}}^t_{\kappa},
 \tilde{\bm{e}}^t_{\lambda} \rangle = \sum
  e^t_{\kappa i} e^t_{\lambda i},
\end{equation}
where $<\; , \;>$ is the inner product in the Euclidean space $X^t$.
This is a Riemannian metric of $X$ pulled-back from $\tilde{X}^t$ to
$X$.  The new squared length of $d{\bm{x}}$ in the input space $X$ due to
the induced metric $g^t_{\kappa \lambda}$ is
\begin{equation}
 \left(ds^2\right)^t = \sum g^t_{\kappa \lambda} dx_{\kappa}
  dx_{\lambda}, 
\end{equation}
which is given by the squared Euclidean length of $d \tilde{\bm{x}}^t$ of layer
$t$.

From (\ref{eq:am2020171124}), we have
\begin{equation}
 \label{eq:am3220171205}
 \langle \tilde{\bm{e}}^t_{\kappa}, \tilde{\bm{e}}^t_{\lambda} \rangle =
 \sum_{i, j, k, l} B^t_{ik}B^t_{jl} \delta_{ij} e^{t-1}_{\kappa l}
 e^{t-1}_{\lambda k},
\end{equation}
where $\delta_{ij}$ is the Kronecker delta.  When $n_t$ is sufficiently large, by the law of large numbers,
\begin{equation}
 \label{eq:am3320171124}
 \frac 1{n_t} \sum_i B^t_{ik} B^t_{il} = {\mathrm{E}}
 \left[ \left\{\varphi' \left(u^t_i \right)\right\}^2 w^t_{ik}w^t_{il}\right].
\end{equation}
If the two terms in the expectation (\ref{eq:am3320171124}) split as
\begin{equation}
 \label{eq:am3420171218}
 {\mathrm{E}} \left[\left\{\varphi' \left( u^t_i \right)\right\}^2 
  w^t_{ij} w^t_{ik}\right] 
  = {\mathrm{E}} \left[ \left\{ \varphi' 
 \left( u^t_i \right)\right\}^2 \right]
  {\mathrm{E}} \left[ w^t_{ij} w^t_{ik}\right], 
\end{equation}
the respective expectations are calculated from
\begin{align}
 \label{eq:am3620180607}
  \chi_{1, t} \left(\sigma^2_t, A^{t-1}\right) 
 & = \sigma^2_t {\mathrm{E}} \left[ \left\{ \varphi' \left(u^t_i \right)\right\}^2
 \right], \\
 \label{eq:am3720180607}
  {\mathrm{E}} \left[w^t_{ik} w^t_{il} \right] & =  \frac{\sigma^2_t}{n_t} \delta_{kl}.
\end{align}
The split lemma given in Appendix II guarantees (\ref{eq:am3420171218}) holds
when $n_t$ is sufficiently large.

\begin{theorem} \upshape
The metric is transformed as
\begin{equation}
 \label{eq:am33}
 g^t_{\kappa \lambda} ({\bm{x}}) = \chi_{1, t} ({\bm{x}}) g^{t-1}_{\kappa \lambda} ({\bm{x}}).
\end{equation}
\end{theorem}

{\it{Proof.}}
From (\ref{eq:am3120180607}) by using
 (\ref{eq:am3220171205})--(\ref{eq:am3720180607}) and
\begin{equation}
 \sum \delta_{kl} e^{t-1}_{\kappa k} e^{t-1}_{\lambda l} =
 \langle \tilde{\bm{e}}^{t-1}_{\kappa},  \tilde{\bm{e}}^{t-1}_{\lambda}
 \rangle 
 = g^{t-1}_{\kappa \lambda},
\end{equation}
we have the theorem.

When a metric tensor is transformed in the following form 
\begin{equation}
 \tilde{g}_{\kappa \lambda}({\bm{x}}) = \rho({\bm{x}})
 g_{\kappa \lambda}({\bm{x}})
\end{equation}
for a scalar function $\rho({\bm{x}})$, the transformation is conformal.  A conformal transformation does not alter the angle of
two tangent vectors, so two orthogonal line elements are always
orthogonal after the transformation.  This implies that the tangent
space is subject to two kinds of transformation without changing the shape:

\begin{description}
 \item[1)] isotropic enlargement/shrinkage;
 \item[2)] rotation.
\end{description}

From (\ref{eq:am33}), the metric induced from layer $t$ is obtained as
\begin{equation}
 g^t_{\kappa \lambda}({\bm{x}}) =
  \chi^{\ast}_t 
 ({\bm{x}}) \delta_{\kappa \lambda},
\end{equation}
where we put
\begin{equation}
 \label{eq:am3920180731}
 \chi^{\ast}_t = \prod^{t-1}_{s=1} \chi_{1, s}.
\end{equation}
When $\sigma^2_t= \sigma^2$ and $t$ is large, $A^t$ converges to
$\bar{A}$.  Hence, we have asymptotically a simple form of metric 
\begin{align}
 g^t_{\kappa \lambda}({\bm{x}}) & \approx \left\{ \bar{\chi}_1 \right\}^{t-1}
 \delta_{\kappa \lambda}, \\
 \bar{\chi}_1 &= \chi_1 \left(\sigma^2, \bar{A}\right),
\end{align}
except for fluctuating terms of order $1/\sqrt{n}$.  When $\varphi$ is the error function, we have
\begin{equation}
 \chi_1(\sigma^2, A) = \frac {\sigma^2}{2 \pi}
 \frac{\sigma^2_b + \sigma^2 A}{\sqrt{1+ 2
  \left(\sigma^2_b + \sigma^2 A \right)}}
\end{equation}
(see Appendix I).

When $\bar{\chi}_1 < 1$, $g^t_{\kappa \lambda}$ converges to 0 because of (\ref{eq:am3920180731}), implying
that $d \tilde{\bm{x}}^t$ shrinks to 0.  Hence no interesting
information processing takes place.  This happens when both $\sigma^2$
and $\sigma^2_b$ are small enough.  When they are large enough,
$\bar{\chi}>1$, and the length of a line element becomes infinitely large as
$t$ becomes large.  However, the signals lie in a bounded region of
$X^t$ because of $\left|x_i \right| \le 1$ except in the case of a ReLU
activation function.  Therefore, if we consider a curve ${\bm{x}}(s)$ in
the input space, its image must be highly curved like a Peano curve,
because the length of two adjacent points is always enlarged (Poole et
al., 2016).  This suggests a chaotic dynamics for ${\bm{x}}^t$.  Such
phenomena are possible when the image $\tilde{X}^t$ of $X$ becomes
highly curved in $X^t$.

We next study the curvature of $\tilde{X}^t$.  When $\sigma^2_t$ are
designed such that $\chi^{\ast}_t \approx 1$, $g^t_{ij}$ neither
diverges nor converges to 0.  So $\tilde{X}^t$ is deformed with slight
expansion or shrinkage but with high curvature.  Interesting information
processing takes place at such an edge of chaos (Yang \& Schoenholtz,
2018).

\section{Curvatures of signal manifolds}

The curvature of the embedded manifold $\tilde{X}^t$ is measured by the
quantities showing how the basis vectors $\tilde{\bm{e}}^t_{\kappa}$ of
the tangent space of $\tilde{X}^t$ change as the point
$\tilde{\bm{x}}^t$ moves in the direction of
$\tilde{\bm{e}}^t_{\lambda}$. (see Figure 6).  
\begin{figure}
 \centering
 \includegraphics[width=7cm]{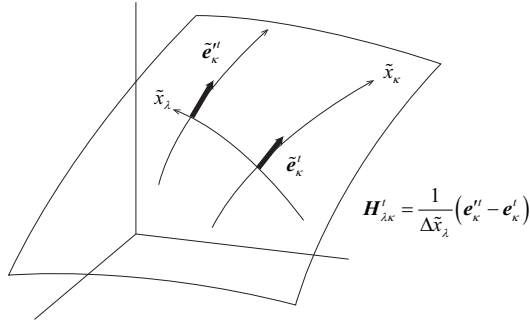}
 \caption{Curvature and affine connections: How $\tilde{\bm{e}}^t_{\kappa}$ changes along the $\tilde{x}^t_{\lambda}$ axis.}
\end{figure} 
By using the directional
derivative $\nabla_{\lambda}$, they are given by the vector
\begin{equation}
 {\bm{H}}^t_{\kappa \lambda} = \nabla_{\lambda} \tilde{\bm{e}}^t_{\kappa},
\end{equation}
where $\nabla_{\lambda}$ is the covariant derivative in the ambient
manifold $X^t$.  However, $X^t$ is a Euclidean space, so
$\nabla_{\lambda}$ is the partial derivative with respect to
$\tilde{x}_{\lambda}$ that is $\nabla_{\lambda} = \partial / \partial x_{\lambda}$.  ${\bm{H}}^t_{\kappa \lambda}$ is a vector in
$X^t$ for fixed indices $\kappa$ and $\lambda$ and is a tensor in $X$
having two indices $\kappa$ and $\lambda$.

We can decompose ${\bm{H}}^t_{\kappa \lambda}$ as a sum of a vector
orthogonal to the tangent space of $\tilde{X}^t \subset X^t$ and a vector in the tangent space.  The orthogonal one shows how $\tilde{X}^t$ is
curved in the orthogonal direction of $\tilde{X}^t$, whereas the
tangential vector demonstrates deformations of
$\left\{\tilde{\bm{e}}^t_{\kappa}\right\}$ inside $\tilde{X}^t$.
Mathematically, the former is the embedding, extrinsic or Euler-Schouten
curvature, showing how $\tilde{X}^t$ is curved in $X^t$.  The latter
represents components of the affine connection that shows how the
coordinate system $\tilde{x}^t_{\kappa}$ is curved inside the manifold
$\tilde{X}^t$.  For the moment, we do not decompose 
${\bm{H}}^t_{\kappa \lambda}$ and simply call it the curvature vector.

The curvature vectors are given in the component form by
\begin{equation}
 H^{t}_{\kappa \lambda i}  =
 \partial_{\lambda} \partial_{\kappa} \varphi 
 \left(u^t_i \right) = \partial_{\lambda} e^t_{\kappa i}, 
\end{equation}
where 
\begin{equation}
 \partial_{\kappa} = 
 \frac{\partial}{\partial x_{\kappa}}.
\end{equation}
We further have, from (\ref{eq:am2020171124}),
\begin{align}
 H^t_{\kappa \lambda i} &= \partial_{\kappa} \left\{
 \sum \varphi' \left(u_i \right) w_{ij} \partial_{\lambda} \tilde{x}^{t-1}_j \right\} \\
 &= \varphi'' \left(u_i \right) \left({\bm{w}}_i \cdot
 \partial_{\kappa} \tilde{\bm{x}}^{t-1}\right) \left(
 {\bm{w}}_i \cdot \partial_{\lambda} \tilde{\bm{x}}^{t-1}\right) \\
  & \mbox{\quad} + \varphi' \left(u_i \right){\bm{w}}_i \cdot
 \partial_{\kappa} \partial_{\lambda} \tilde{\bm{x}}^{t-1},
\end{align}
where ${\bm{w}}_i = \left(w^t_{ij}\right)$ and we omit the superscript
$t$ attached to $w_{ij}$ and $u_i$.  By noting
\begin{equation}
 {\bm{H}}^{t-1}_{\kappa \lambda} = \partial_{\kappa} \partial_{\lambda}
  \tilde{\bm{x}}^{t-1},
\end{equation}
the following recursive equation is obtained
\begin{equation}
 H^t_{\kappa \lambda i} = \varphi'' \left(u_i \right)
 \left({\bm{w}}_i \cdot \tilde{\bm{e}}^{t-1}_{\kappa}\right)
 \left({\bm{w}}_i \cdot \tilde{\bm{e}}^{t-1}_{\lambda}\right)
 + \varphi' \left(u_i \right){\bm{w}}_i \cdot
 {\bm{H}}^{t-1}_{\kappa \lambda}.
\end{equation}

We define the magnitude of the curvature vector by
\begin{equation}
 \left|{\bm{H}}^t_{\kappa \lambda}\right|^2 =
  \langle {\bm{H}}^t_{\kappa \lambda}, {\bm{H}}^t_{\kappa \lambda}
 \rangle =  \sum_i
 \left(H^t_{\kappa \lambda i}\right)^2
 \end{equation}
and replace the summation by the expectation by using the law of large
numbers.  Then, we have
\begin{align}
 \left|{\bm{H}}^t_{\kappa \lambda}\right|^2 &=
  n_t  {\mathrm{E}} \left[ \left\{ \varphi''(u_i) \right\}^2 \left({\bm{w}} \cdot
 \tilde{\bm{e}}^{t-1}_{\kappa} \right)^2 \left({\bm{w}} \cdot
 \tilde{\bm{e}}^{t-1}_{\lambda}\right)^2  \right] \\
 & \quad + 2 n_t {\mathrm{E}} \left[ \varphi'(u_i) \varphi''(u_i)
 \left({\bm{w}} \cdot \tilde{\bm{e}}^{t-1}_{\kappa}\right)
 \left({\bm{w}} \cdot \tilde{\bm{e}}^{t-1}_{\lambda}\right)
 \left({\bm{w}}\cdot
 \partial_{\kappa} \tilde{\bm{e}}^{t-1}_{\lambda}\right)\right] \\
 & \qquad + n_t {\mathrm{E}} \left[\left\{ \varphi'(u_i)\right\}^2
 \left({\bm{w}} \cdot {\bm{H}}^{t-1}_{\kappa \lambda}\right)^2 \right].
\end{align}
It consists of three terms.  The second term vanishes because it is an
odd function of ${\bm{w}}$.  We use the split lemma given Appendix II that the terms of functions of $u_i$ and the remaining terms (the second or fourth order
terms of ${\bm{w}}$) split in the expectation.  This is because the terms
$\varphi'(u_i)$ and $\varphi''(u_i)$ have the self-averaging property
(the mean field approximation).

The first term is the product of
\begin{equation}
  {\mathrm{E}} \left[ \left\{ \varphi''(u^t_i) \right\}^2 \right]
  = \frac 1{\sigma^4_t} \chi_2(\sigma^2_t, A^{t-1}) 
\end{equation}
and
\begin{equation}
 {\mathrm{E}}
 \left[ \left({\bm{w}} \cdot \tilde{\bm{e}}^{t-1}_{\kappa}\right)^2
 \left({\bm{w}} \cdot \tilde{\bm{e}}^{t-1}_{\lambda}\right)^2\right] =
 \frac{\sigma^4_t}{n^2_t} \left(1+2 \delta_{\kappa \lambda}\right)
 \left|\tilde{\bm{e}}^{t-1}_{\kappa} \right|^2
  \left| \tilde{\bm{e}}^{t-1}_{\lambda} \right|^2,
\end{equation}
where we use
\begin{equation}
 E \left[ w_i w_j w_k w_l \right] = 
 \frac{\sigma^4_t}{n^2_t} \left( \delta_{ij} \delta_{kl}+
 \delta_{ik} \delta_{jl}+ \delta_{il} \delta_{jk}\right).
\end{equation}
The third term is
\begin{equation}
 {\mathrm{E}} \left[ \varphi'^{2}\right]{\mathrm{E}}
 \left[ \left({\bm{w}} \cdot {\bm{H}}^{t-1}_{\kappa \lambda}\right)^2
 \right] = \chi_1 \left(\sigma^2_t, A^{t-1}\right)
 \left|{\bm{H}}^{t-1}_{\kappa \lambda}\right|^2.
\end{equation}
Therefore, we have 
\begin{equation}
 \left|{\bm{H}}^t_{\kappa \lambda} \right|^2 = \chi_1
\left(\sigma^2_t, A^{t-1}\right) \left|{\bm{H}}^{t-1}_{\kappa \lambda} \right|^2
 + \frac 1{n_t} \left( 1+ 2 \delta_{\kappa \lambda}\right) \left\{\chi_2
\left(\sigma^2_t, A^{t-1}\right)\right\}
 \left(g^{t-1}_{\kappa \kappa}\right)^2
 \left(g^{t-1}_{\lambda \lambda}\right)^2.
\end{equation}

We did not distinguish the embedding curvature and affine connection.
Since ${\bm{w}}_i$ is isotropically distributed in $X^t$, when
$n^{\perp}_t$ is the number of dimensions orthogonal to $\tilde{X}^t$,
$\left(n^{\perp}_t/n_t\right)$ $\left|{\bm{H}}_{\kappa
\lambda}\right|^2$ is the extrinsic curvature tensor and
$\left(1-n^{\perp}_t / n_t \right)$ $\left|{\bm{H}}_{\kappa
\lambda}\right|^2$ is due to the affine connection, because of the
equipartition property.

We further simplify the situation by defining the scalar curvature
\begin{equation}
 \gamma^2_t = \frac 1{n_t} \sum \left(g^{-1}_t \right)^{\kappa \mu}
 \left(g^{-1}_t \right)^{\lambda \nu}
 H^t_{\kappa \lambda i} H^t_{\mu \nu j} \delta_{ij},
\end{equation}
where $\left(g^{-1}_t \right)^{\kappa \lambda}$ is the inverse matrix
of $\left(g^t_{\kappa \lambda}\right)$.

From
\begin{equation}
 \left(g^{-1}_t \right)^{\kappa \lambda} = \frac 1{\chi_{1, t-1}}
 \left(g^{-1}_{t-1}\right)^{\kappa \lambda}
\end{equation}
we have the following recursive equation
\begin{equation}
 \label{eq:am6220171212}
 \gamma^2_t = \frac{\gamma^2_{t-1}}{\chi_{1, t-1}} + 3 \frac{\chi_{2,
 t-1}}{n_t \left(\chi_{1, t-1}\right)^2}.
\end{equation}
The first term on the right-hand side of (\ref{eq:am6220171212}) shows
the curvature inherited from the input with decay factor $\chi^{-1}_1$
and the second term is the newly created curvature in the layer.  This
gives
\begin{equation}
 \gamma^2_t = 3 \frac 1{n_t} \sum^{t-1}_{s=1}
 \frac{\chi_{2, t-s}}{\chi^2_{1, t-s}} \left(
 \prod^s_{r=1} \frac 1{\chi_{1, t-r}}\right).
\end{equation} 
When $t$ is sufficiently large and $\sigma^2_t = \sigma^2$ and
$\chi_1 = \chi_{1, t} >1$, we have
\begin{equation}
 \gamma^2_t = \frac{3 \chi_2}{n_t \chi_1 \left(\chi_1-1 \right)},
\end{equation}
implying that the scalar curvature converges to a small constant.  This
is the result obtained by Poole et al. (2016) when $X$ is a
1-dimensional curve.  When $\chi_1<1$, $\gamma^2_t$ diverges to
infinity, provided $n_t$ are finite although they are large, in spite that $\tilde{X}^t$ shrinks.

When $\chi^{\ast}_t \approx 1$, which implies the network dynamics is at
the edge of chaos (Poole et al. 2016; Schoenholz et al. 2017), the
induced metric is
\begin{equation}
 g^t_{\kappa \lambda} \approx \delta_{\kappa \lambda},
\end{equation}
except for higher-order fluctuations, keeping the length and orthogonality.  However, even though new creations of curvature are small (the second terms of (\ref{eq:am6220171212}) is of order $1/n_s$), they accumulate and
\begin{equation}
 \label{eq:am6720180618}
 \gamma^2_t \approx 3 \sum^{t-1}_{s=1} \frac 1{n_s}
 \frac{\chi_{2, s}}{\chi^2_{1, s}}
\end{equation}
diverges to infinity, provided $n_s$ are finite.  Hence, although the
metric is well controlled, $\tilde{X}^t$ is highly deformed since the
coordinates $\tilde{\bm{x}}^t$ are highly distorted in $\tilde{X}^t$.

\section{Law of distance}

Let ${\bm{x}}^{t-1}$ and ${\bm{y}}^{t-1}$ be two input signals at layer
$t$.  Their outputs are ${\bm{x}}^t$ and ${\bm{y}}^t$, respectively.
Let
\begin{equation}
 D_t = D \left({\bm{x}}^t, {\bm{y}}^t \right) =
 \frac 1{n_t} \sum \left( x^t_i-y^t_i \right)^2
\end{equation}
be the squared Euclidean distance between two signals ${\bm{x}}^t$ and
${\bm{y}}^t$ divided by $n_t$.  We study how $D_t$ is related to
$D_{t-1}$.  When ${\bm{x}}^{t-1}$ and ${\bm{y}}^{t-1}$ are infinitesimally
close, $d{\bm{x}}^{t-1}={\bm{x}}^{t-1}-{\bm{y}}^{t-1}$ belongs to the
tangent space and it expands or shrinks by a scalar factor $\chi_1$
depending on whether it is larger than 1 or not.  Here, we study how the
distance develops when ${\bm{y}}^{t-1}$ and ${\bm{x}}^{t-1}$ are not
necessarily close.

It is easier to study how the overlap
\begin{equation}
 C_t=C \left({\bm{x}}, {\bm{y}} \right) =
 \frac 1{n_t \sqrt{A \left({\bm{x}}^t\right)A \left({\bm{y}}^t\right)}}
  \sum x^t_i y^t_i
\end{equation}
develops.  The distance and overlap are related by
\begin{equation}
 \label{eq:am7120180601}
 D \left({\bm{x}}, {\bm{y}} \right) = A({\bm{x}}) +
 A \left({\bm{y}} \right) - 2 \sqrt{A({\bm{x}})A({\bm{y}})}C \left({\bm{x}},
    {\bm{y}}\right), 
\end{equation}
so when we know the law of overlap
\begin{equation}
 C_t = \psi \left(C_{t-1}\right),
\end{equation}
the law of distance is obtained in the form
\begin{equation}
 \label{eq:am7220180618}
 D_t = \xi \left(D_{t-1}\right)
\end{equation}
by using (\ref{eq:am7120180601}).  For simplicity, we assume
that $A({\bm{x}})$ and $A({\bm{y}})$ are equal to $A^t$.  Then,
\begin{equation}
 D_t = 2 A^t \left(1-C_t \right)
\end{equation}
so we have the explicit form of $\xi(D)$ as
\begin{equation}
  \xi \left(D_{t-1}\right) = 2 A^t \left\{ 1-\psi
 \left(1- \frac{D_{t-1}}{2 A^t}\right)\right\}.
\end{equation}

Two random variables
\begin{align}
 u &= \sum w_j x_j + b_j, \\
 u' &= \sum w_j y_j + b_j,
\end{align}
in which indices $i$ and $t$ are omitted for simplicity, are jointly
Gaussian with mean 0 and their variances and covariances are written as
\begin{align}
 \label{eq:am7020171130}
 E \left[ u^2 \right] &= E \left[ u'^2 \right] = \sigma^2_{A^t}, \\
 \label{eq:am7120171130}
 E \left[u u' \right] &= \sigma^2 A^t C \left({\bm{x}}, {\bm{y}}
 \right) + \sigma^2_b = \sigma^2_{A^t C}.
\end{align}
Then, we have the law of overlap from
\begin{equation}
 C = \frac 1{A^t} E \left[ \varphi(u) \varphi
 \left(u'\right)\right] 
\end{equation}
and by tedious calculations given in Appendix III, 
\begin{equation}
 \label{eq:am8020180618}
 C_t = \frac 1{2 \pi A^t} \cos^{-1} \left(
 -\frac{C_{t-1} A^t \sigma^2_t + \sigma^2_b}{\sigma^2_{A^t} 
 \sqrt{1+\sigma^2_{A^t}}} \right).
\end{equation}

This gives the law of distance $\xi(D)$.  It satisfies
\begin{equation}
 \label{eq:am8220180607}
 \xi(0) = 0
\end{equation}
and is a monotonically increasing function of $D$.  When $\sigma^2_t =
\sigma^2$, we obtain
\begin{equation}
 \xi'(0) = \bar{\chi}_1
\end{equation}
from the direct calculation or from the fact that the length of
$d{\bm{x}}$ increases by the factor $\bar{\chi}_1$.  Hence when
$\bar{\chi}_1 <1$, the distance converges to 0, so information
processing is poor.  When $\bar{\chi}_1>1$, the dynamics of distance
(\ref{eq:am7220180618}) has a solution other than the 0 specified in equation
(\ref{eq:am8220180607}),
\begin{equation}
 \bar{D} = \xi \left( \bar{D}\right),
\end{equation}
which is uniquely determined.  See Fig. 7.  
\begin{figure}
 \centering
 \includegraphics[width=7cm]{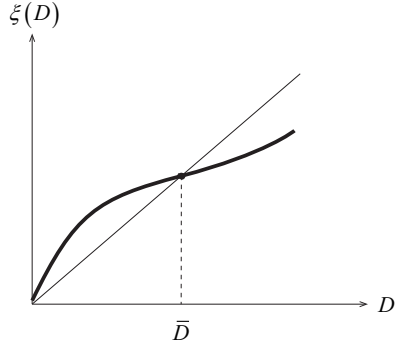}
 \caption{Law of distance}
\end{figure} 
The dynamics of distance
converges to $\bar{D}$ as $t \rightarrow \infty$, since the other
equilibrium given in (\ref{eq:am8220180607}) is unstable.  This implies
that the distance between any two input signals ${\bm{x}}_1$ and
${\bm{x}}_2 \; \left({\bm{x}}_1 \ne {\bm{x}}_2 \right)$ converges to
$\bar{D}$ as $t$ goes to infinity, provided $n_t$ are sufficiently
large.

However, this is problematic, because $X$ is an $n$-dimensional continuum.
The neural transformation specified in equations (\ref{eq::am520180601}) and
(\ref{eq::am620180601}) is continuous, so it is impossible that the
distances of all pairs of two input signals become $\bar{D}$.  A
set $Z$ of points in which any two points have the same distance
$\bar{D}$ consists of at most $N+1$ points in an $N$-dimensional space.
Therefore, it is impossible that $X$ is mapped on $Z$ by a continuous
transformation.  There exist frustrations between the distance and
continuity, caused by the fimiteness of $n_t$.

Let $\xi^{\ast t}$ be the $t$ times concatenation of the distance law:
\begin{equation}
 \xi^{\ast t}(D) = \xi \left\{ \xi \cdots \left\{\xi(D)\right\} \cdots
		    \right\}. 
\end{equation}
It converges to the step function
\begin{equation}
 \label{eq:am8720180604}
 {\mathop{\lim}_{t \rightarrow \infty}} \xi^{\ast t}(D) =
 \left\{
  \begin{array}{ll}
  \bar{D}, & D>0, \\
  0, & \mathrm{otherwise}.
 \end{array}
 \right.
\end{equation}
as $t$ goes to infinity, where we assume $n \rightarrow \infty$.  We
consider a straight line
\begin{equation}
 {\bm{x}}(s) = s {\bm{x}}
\end{equation}
in $X$ as a simple example and let
\begin{equation}
 D(s) = D \left[{\bm{x}}(0), {\bm{x}}(s)\right],
\end{equation}
which is the squared distance from the origin ${\bm{x}}(s)=0$ to
${\bm{x}}(s)$.  Then, the distance between two points after the $t$-th layer
is given by
\begin{equation}
 D^t(s) = \xi^{\ast t}\left\{D(s)\right\}.
\end{equation}
When we consider the distance between ${\bm{x}} \left(s_1 \right)$ and
${\bm{x}} \left(s_2 \right)$, the distance after the transformation is
given by
\begin{equation}
 D \left[ \tilde{\bm{x}}^t \left(s_1 \right), \tilde{\bm{x}}^t
    \left(s_2 \right)\right] = \xi^{\ast t}(s_2-s_1).
\end{equation}
From (\ref{eq:am8720180604}), we see that the curve
$\tilde{\bm{x}}^t(s)$ has the fractal-like structure since we have $\xi^{\ast
t}(\alpha s) = \xi^{\ast t}(s)$ for any $\alpha$ when $t$ is large.
There exist adversarial examples in a deep network. They are due to the fractal nature of the transformation.

The convergence of distance to $\bar{D}$ is derived in the situation
when $n \rightarrow \infty$ and then $t \rightarrow \infty$.  It is
known that
\begin{equation}
 \mathop{\lim}_{t \rightarrow  \infty} \mathop{\lim}_{n \rightarrow
  \infty}
  \xi^{\ast t}(D) = \mathop{\lim}_{n \rightarrow \infty} \mathop{\lim}_{t \rightarrow
  \infty} \xi^{\ast t}(D)
\end{equation}
does not necessarily holds (see Amari, Yoshida \& Kanatani, 1977, for the case
of random recurrent networks).  In reality, both $n$ and $t$ are finite.
When $n$ is finite, $\xi(D)$ suffers from random fluctuations, so
(\ref{eq:am8720180604}) does not hold exactly.  Hence the image of a
curve ${\bm{x}}(s)$ is still a continuous curve, and the distance
between ${\bm{x}}(s)$ and ${\bm{x}}(s')$, $s \ne s'$, is not exactly to
$\bar{D}$.  It is outside of the scope of the present papaer but is interesting to study how $D({\bm{x}}(0), {\bm{x}}(s))$
behaves for large but finite $n$ as $t$ increases.  The non-uniform
convergence with respect to $n$ and $t$ is also seen in (\ref{eq:am6720180618}).

\section{Collapse of manifolds when $n_t < n_{t-1}$}

We have discussed the case with $n_t \ge n_{t-1}$, where the dimension
numbers are non-decreasing.  When $n_t < n_{t-1}$, dimension reduction
takes place and the situation is completely different.  In this
case, there exists the null space of weight matrix ${\bm{W}}^t= \left(w^t_{ij} \right)$,
\begin{equation}
 N = \left\{{\bm{x}} \left| \; {\bm{W}}^t {\bm{x}}=0 \right\} \right..
\end{equation}
In other words, for any ${\bm{x}}$ and ${\bm{n}} \in N$,
\begin{equation}
 \bm{W} {\bm{x}} =  {\bm{W}} \left({\bm{x}} + {\bm{n}}\right).
\end{equation}
This means that the null directions are collapsed by multiplication by ${\bm{W}}^t$. 

The image $\tilde{X}^{t-1}$ is a highly curved submanifold in $X^{t-1}$,
so this dimension reduction tears $\tilde{X}^{t-1}$ and many points in
$\tilde{X}^{t-1}$ are mapped to a point in $X^t$ (see Figure 8).  
\begin{figure}
 \centering
 \includegraphics[width=7cm]{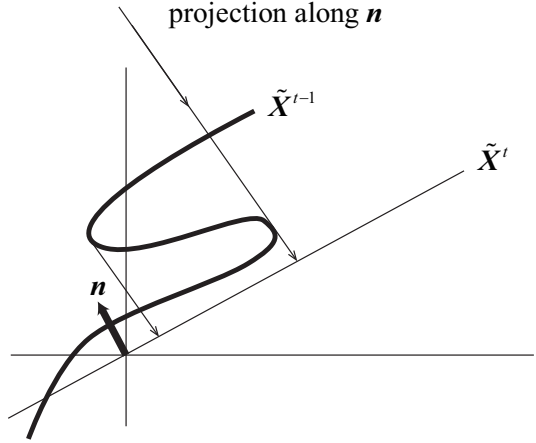}
 \caption{Dimension reduction and tearing of $\tilde{\bm{X}}^{t-1}$}
\end{figure} 
Then,
nonlinear $\varphi$ is applied. The activity, metric and curvature
develops in the same way after the dimension reduction. How is the structure of $\tilde{X}^t$ after dimension reduction is an interesting problem to be studied further.

\section{Conclusions}

Using the statistical neurodynamics of a multilayer perceptron of random connections, we have studied how the signal space geometry, metric, curvature, and distance between signals develop as signals are processed in deep networks. The present asymptotic theory assumes that the number $n_t$ of neuros in each layer is {\it{sufficiently}} large.  We then found that the metric is conformally mapped from layer to layer,
where the scale factor $\chi_1$ plays a fundamental role in this mapping.  When
$\chi^{\ast}_t \approx 1$, the dynamics of $\tilde{\bm{x}}^t$ is
chaotic and rich information processing takes place (Yang and Schonholtz, 2017).  The curvature
tensor and scalar curvature are obtained explicitly, showing that the curvature
diverges to infinity when $\chi^{\ast}_t =1$ provided $n$ is finite.  That is, creations of new curvatures are of order $1/n$, infinitesimally small, but are accumulated to infinity.  This implies that the finite $n$ effect is important.  The metric $g^t_{\kappa \lambda}$ also fluctuates from $\chi^{\ast}_t \delta_{\kappa \lambda}$ and accumulates, when $n$ is finite, which is important for accumulation of curvature.  How the distance between two input signals develops is also shown.

Our results hold in the limit of $t \rightarrow \infty$ and $n \rightarrow \infty $, but in reality $t$ and $n$ are finite.  So we need to
study the effect of finiteness, which resolves apparently embarrassing results of the present theory. We need to study the finiteness effect
carefully to resolve the contradiction. We have mostly focused on the case of $n_t \ge n_{t-1}$.  However, information reduction takes place when $n_t < n_{t-1}$. We need to study this case in details in future work.

\section*{Appendix I: Useful formulas}

The explicit forms of the fundamental functions $\chi_p (A)$ are obtained
from the following formulas, where $\varphi$ is the error function.
\begin{description}
 \item[1)]
\begin{equation}
 \label{eq:am9420180607}
 \int^{\infty}_{-\infty} \varphi (au) \varphi (bu) Du =
 \frac 1{2 \pi} \cos^{-1} 
 \left( \frac{-ab}{\sqrt{1+a^2}\sqrt{1+b^2}}\right),
\end{equation}
 \item[2)] 
\begin{equation}
 \label{eq:am9520180607}
 \int^{\infty}_{-\infty} \left\{ \varphi'(au)\right\}^2 Du =
 \frac 1{2 \pi}　\frac{a^2}{ \sqrt{1+2a^2}},
\end{equation}
 \item[3)]
\begin{equation}
 \label{eq:am9620180607}
 \int^{\infty}_{-\infty} \left\{ \varphi'' (au)\right\}^2 Du =
 \frac 1{2 \pi} \left (\frac{a^2}{\sqrt{1+2a^2}} \right )^3.
\end{equation}
\end{description}

By putting $a=b= \sigma_A$ in 1), we have
\begin{equation}
 \chi_0(A) = \frac 1{2 \pi} \cos^{-1} \left\{
 \frac 1{1+\sigma^2_b + \sigma^2 A} -1 \right\}.
\end{equation}
By putting $a= \sigma_A$ in 2) and 3), we have
\begin{align}
 \chi_1(A) &= \frac {\sigma^2}{2 \pi} \frac{\sigma^2_b + \sigma^2 A}{\sqrt{1+2
 \left( \sigma^2_b + \sigma^2 A\right)}}, \\
 \chi_2(A) &= \frac {\sigma^4}{2 \pi}  \left( \frac{\sigma^2_b + \sigma^2 A}{\sqrt{1+2
 \left( \sigma^2_b + \sigma^2 A\right)}}  \right)^3.
\end{align}

Equations (\ref{eq:am9520180607}) and (\ref{eq:am9620180607}) are easy
to prove, because $\varphi'$ is standard Gaussian.  We prove 1).  We
have
\begin{align}
 \int^{\infty}_{-\infty} \varphi(au) \varphi(bu) Du &=
 \frac 1{(\sqrt{2 \pi})^3} \int^{\infty}_{-\infty}
 du e^{-\frac{u^2}2} \int^{au}_{-\infty} e^{-\frac{v^2}2} 
 dv \int^{bu}_{-\infty} e^{-\frac{w^2}2} dv \\
 &= \frac 1{(\sqrt{2 \pi})^3} \int_R \exp 
 \left\{ -\frac{u^2+v^2+w^2}2 \right\} dudvdw,
\end{align}
where $R$ is the region
\begin{equation}
 R = \left\{ (u, v, w) \in S_1 \cap S_2 \right\},
\end{equation}
\begin{align}
 S_1 &: v \le au, \\
 S_2 &: w \le bu.
\end{align}
$S_1$ and $S_2$ are the regions under the two planes $v=au$ and $w=bu$, both
passing through the origin.  The normal vectors of the two surfaces are
\begin{equation}
 {\bm{n}}_1 = (\frac a{\sqrt{1+a^2}},\frac {-1}{\sqrt{1+a^2}},0), \quad
 {\bm{n}}_2 = (\frac {b}{\sqrt{1+b^2}},0,\frac {-1}{\sqrt{1+b^2}}).
\end{equation}
The angle between $S_1$ and $S_2$ is $\theta= \pi-\cos^{-1}
\left({\bm{n}}_1 \cdot {\bm{n}}_2 \right) = \cos^{-1} \left(-{\bm{n}}_1
\cdot {\bm{n}}_2 \right)$.  Hence, we have
\begin{equation}
 1) = \frac 1{2 \pi} \cos^{-1} \frac{-ab}{\sqrt{1+a^2}\sqrt{1+b^2}}. 
\end{equation}
By putting $a=\sigma_A$,
\begin{equation}
 \chi_0(A) = \frac 1{2 \pi} \cos^{-1}
 \frac{-\left(\sigma^2_A+ \sigma^2_b\right)}{1+\sigma^2_A+ \sigma^2_b}. 
\end{equation}
This is monotonically increasing in $0 \le A \le 1$.

\section*{Appendix II}

{\textbf{Split Lemma}} \quad When $w_1, \cdots, w_n$ are independent
random variables subject to $N \left(0, \sigma^2/n \right)$, the
expectation of the product of two terms splits as
\begin{equation}
 {\mathrm{E}} \left[ f({\bm{w}} \cdot {\bm{x}}) k \left({\bm{w}} \right)\right] =
 {\mathrm{E}} \left[ f({\bm{w}} \cdot {\bm{x}})\right] E \left[ k \left({\bm{w}}
 \right)\right] 
\end{equation}
for arbitrary analytical functions $f$ and $k \left({\bm{w}} \right)=
w_i w_j$ or $w_i w_j w_k w_l$, etc., when $n$ is sufficiently large.

\begin{proof}\upshape
We prove the case of $k({\bm{w}})= w_i w_j$.  For
\begin{equation}
 u = {\bm{w}} \cdot {\bm{x}} = \sum w_i u_i
\end{equation}
we define
\begin{equation}
 \check{u}_{ij} = u-w_i x_i - w_j x_j.
\end{equation}
Then, by Taylor expansion, we have
\begin{equation}
 f(u) = f \left(\hat{u}_j \right) + f' \left(\hat{u}_{ij}\right)
 \left(w_i x_i + w_j x_j \right)
\end{equation}
since $w_i$ and $w_j$ are small.  Hence,
\begin{align}
 {\mathrm{E}} \left[f(u)k \left({\bm{w}} \right)\right] &= 
 {\mathrm{E}} \left[ f \left(\hat{u}_{ij} \right)\right] + \mbox{higher-order terms} \\
 &= {\mathrm{E}} \left[ f \left(\hat{u}_{ij}\right)\right] 
 {\mathrm{E}} \left[k \left({\bm{w}} \right)\right]+ \mbox{higher-order terms}
\end{align}
since $\hat{u}_{ij}$ and $k \left({\bm{w}} \right)$ are independent.  We
again have
\begin{equation}
 {\mathrm{E}} \left[ f \left(\hat{u}_{ij} \right)\right] =
 {\mathrm{E}} \left[f(u)\right] + \mbox{higher-order terms}.
\end{equation}
\end{proof}
The higher-order terms vanish as $n \rightarrow \infty$, so we have the
lemma.  The proof is similar when $k({\bm{w}})= w_i w_j w_k w_l$ .

\section*{Appendix III: Law of distance}

Let $\varepsilon$ and $\nu$ be two independent standard Gaussian random
variables subject to $N(0, 1)$.  From equations (\ref{eq:am7020171130}) and
(\ref{eq:am7120171130}), we have new representations of $u$ and  $u'$:

\begin{align}
 u &= \sigma_{A} \varepsilon, \\
 u' &= \alpha \varepsilon + \beta \nu,
\end{align}
where 
\begin{equation}
 \alpha = \frac{\sigma^2_{AC}}{\sigma_A}, \quad
 \beta = \frac{\sqrt{\sigma^4_A-\sigma^4_{AC}}}{\sigma_A}.
\end{equation}
We see that
\begin{equation}
 E \left[ \varphi(u) \varphi \left(u' \right)\right] =
 \frac 1{(2 \pi)^2} \int e^{-\frac{\varepsilon^2+\nu^2}2}
 \int^{\sigma_A{\varepsilon}}_{-\infty}
 e^{-\frac{x^2}2} \int^{\alpha \varepsilon+ \beta \nu}_{-\infty}
 e^{-\frac{y^2}2} dxdyd \varepsilon d \nu.
\end{equation}
We first calculate the integration by $y$ and $\nu$, where we use
\begin{equation}
 \int^{\infty}_{-\infty} \varphi (\alpha \varepsilon+ \beta \nu)
 D \nu = \varphi \left(\frac{\alpha \varepsilon}{\sqrt{1+
  \beta^2}}\right). 
\end{equation}
Then
\begin{equation}
 {\mathrm{E}} \left[ \varphi(u) \varphi \left(u' \right)\right] =
 {\mathrm{E}} \left[ \varphi \left(\sigma_A \varepsilon \right)
 \varphi \left( \frac{\alpha \varepsilon}{\sqrt{1+ \beta^2}}\right)
 \right].
\end{equation}
Since this is 
\begin{equation}
 \frac 1{2 \pi} \int \int_R \exp \left\{
 -\frac{y^2 + \nu^2}2 \right\} dy d \nu
\end{equation}
in the region
\begin{equation}
 R: y \le \alpha \varepsilon + \beta \nu,
\end{equation}
calculating carefully, we finally have equation (\ref{eq:am8020180618}).

\end{document}